\documentclass[10pt,conference]{IEEEtran}
\usepackage{times}
\usepackage[latin1]{inputenc}
\usepackage[hyphens]{url}
\usepackage{hyperref}
\usepackage{color}
\usepackage{graphicx}
\usepackage{amsmath}
\usepackage[noadjust]{cite}
\usepackage[font=small,justification=centering,labelsep=period]{caption}

\begin{document}
\bstctlcite{IEEEexample:BSTcontrol}

\title{\textbf{\Large Multiclass Disease Predictions Based on Integrated Clinical and Genomics Datasets} \\
}

\author{\IEEEauthorblockN{\large Moeez M. Subhani}
\IEEEauthorblockA{College of Engineering and Technology}
University of Derby\\
Derby, England \\
Email: {\tt m.subhani@derby.ac.uk}
\and
\IEEEauthorblockN{\large Ashiq Anjum}
\IEEEauthorblockA{College of Engineering and Technology \\
University of Derby\\
Derby, England \\
Email: \tt a.anjum@derby.ac.uk}
}

\maketitle

\begin{abstract}
Clinical predictions using clinical data by computational methods are common in bioinformatics. However, clinical predictions using information from genomics datasets as well is not a frequently observed phenomenon in research. Precision medicine research requires information from all available datasets to provide intelligent clinical solutions. In this paper, we have attempted to create a prediction model which uses information from both clinical and genomics datasets. We have demonstrated multiclass disease predictions based on combined clinical and genomics datasets using machine learning methods. We have created an integrated dataset, using a clinical (ClinVar) and a genomics (gene expression) dataset, and trained it using instance-based learner to predict clinical diseases. We have used an innovative but simple way for multiclass classification, where the number of output classes is as high as 75. We have used Principal Component Analysis for feature selection. The classifier predicted diseases with 73\% accuracy on the integrated dataset. The results were consistent and competent when compared with other classification models. The results show that genomics information can be reliably included in datasets for clinical predictions and it can prove to be valuable in clinical diagnostics and precision medicine. \\ 
\end{abstract}

\begin{IEEEkeywords}
Clinical; Genomics; Data Integration; Machine Learning; Disease Prediction; Classification; Bioinformatics.
\end{IEEEkeywords}

\section{Introduction}
The medical science is rich with various types of datasets ranging from clinical to genomics datasets. The clinical datasets are diverse in terms of their nature, format and the information they contain. On the other hand, genomics datasets are intrinsically enormous in size and dimensions, and so is the information contained in them \cite{subhani2016}. The genomic information can be considered as the backbone of clinical information since the genomic structure derives the physical characteristics of any organism. If the two pieces of information are connected, it may help to improve the overall medical research by finding more accurate and advanced clinical diagnostic solutions. The connection essentially means to integrate clinical and genomics datasets. This is also a way forward in precision medicine studies, where medical practitioners want to make clinical decisions based on both clinical and genomics parameters and not just one of them \cite{higdon2015}\cite{kim2014}.

However, the research to establish or explore this connection is not very commonly sought in the state-of-the-art \cite{subhani2016}\cite{hamid2009,louie2007,ritchie2015}. The datasets from clinical and genomics sources are mainly used independently in their respective research domains. From literature review, it has been observed that most clinical prediction studies have been limited to either clinical datasets \cite{delen2005,fridley2012,kim2013,lasko2013,palaniappan2014,raikwal2012,mcclatchey2013} or genomics datasets \cite{akavia2010,brown2000,kim2012,lu2005,schadt2005,zhu2008,yang2014}. One common factor among these studies is that almost all of them are prediction studies, which establishes the fact that the trend for clinical predictions has long prevailed in research. 

Although there are some studies which have attempted towards the inter-domain research, the trend does not seem to be very progressive. For example, \cite{nevins2003} used decision trees to predict breast cancer outcomes. Similarly, \cite{lee2009} employed multiple regression and statistical methods to infer associations, and \cite{kim2014} used a graph-based approach to predict cancer clinical outcomes from multi-omics data. All these studies used integrated datasets for prediction or association studies using various approaches. However, most of these approaches are now outdated due to limitations in terms of their performance or accuracy \cite{nevins2003}\cite{lee2009}. The approach in \cite{kim2014} (combination of regression, Bayesian networks, and evolutionary neural networks) is more advanced and promising but this study is limited to binary classifications and multi-omics data only \cite{kim2015}\cite{kim2017}.

The research work mentioned above show that prediction based studies are common in the literature. The most popular or commonly sought predicting factors are survival rate and disease recurrence rate. However, we could not find any disease prediction model in the literature based on combined clinical and genomics data information. A typical disease prediction model, as we define, takes information from both clinical and genomics datasets and predicts disease(s) in a patient. This can be achieved when we have both clinical and genomics datasets available for a variety of diseases. Hence, we are attempting to design a disease prediction model which aims to predict possible medical condition(s) in a patient using information from both clinical and genomics datasets.

\begin{table*}[htbp]
\caption{CLINVAR DATASET.}
\label{table 1}
\centering
\begin{small}
\begin{sc}
\begin{tabular}{p{0.115\linewidth}p{0.115\linewidth}p{0.115\linewidth}p{0.115\linewidth}p{0.115\linewidth}p{0.115\linewidth}p{0.115\linewidth}}
\hline
Gene & Condition & Clinical Significance & Chromosome No. & Location & Variation ID & Allele ID  \\
\hline
AKAP & Long QT syndrome & Benign/Likely benign &7 & 92001306 & 136347 & 140050  \\
AKT2 & Colorectal Neoplasms & Likely pathogenic & 19 & 40236313 & 376039 & 362918 \\
APC & Hereditary cancer-predisposing syndrome &	Pathogenic & 5 & N/A & 181836 &	181126 \\
... & ... & ... & ... & ... & ... & ... \\
\hline
\end{tabular}
\end{sc}
\end{small}
\end{table*}

From ClinVar and Expression Atlas databases, we have been able to construct such dataset which contains both clinical parameters as well as gene expression values in a single dataset for several patients. Since the data retrieved from these databases is in eXtensible Markup Language (XML) format, we can create a very flexible schema for this dataset. Using this dataset, we can train a model to learn the diseases in various subjects. As an initial attempt to prove the concept, we have used the k-Nearest Neighbours (kNN) algorithm for the learning model, which is an instance based learner \cite{duda1973}. Considering the size and complexity of the dataset, kNN appears to be a reasonable choice of learning method since it learns the classification function only locally. 

Genomics based clinical diagnosis does not exist in clinical environments. Traditionally, disease predictions are made using regular clinical practices only. Our disease prediction model can provide a genomic signature to verify the disease existence or possible occurrence. Hence, this model not only will help the medical practitioners to gain another step of confidence in terms of clinical diagnosis, but also help advance the precision medicine research.

The rest of the paper is arranged as follows. Section II discusses the challenges for data integration. Section III explains the data integration model. Section IV gives details of the prediction model and the algorithm along with the implementation details. Section V presents the results, followed by discussion in Section VI and conclusion in Section VII.

\section{Clinical and Genomics Data Integration Challenges}

The integration of clinical and genomics datasets is crucial to move towards precision medicine. The medical conditions of each person are transcribed from the underlying genomics structure. Hence, it is critical to bring forward the genomic information to play part in the clinical diagnostics \cite{higdon2015}\cite{kim2014}. The main challenge is to find a way to integrate datasets which are completely different from each other in terms of their nature, size, and properties. 

Most biological databases have standardised the data storage in XML formats. European Molecular Biology Laboratory (EMBL) took an initiative in 2000 to provide access of all the flat files data in XML format \cite{robinson2002}. XML provides more flexibility in terms of storage, transport and integration of complex biological datasets \cite{lacroix2002}. The format also provides the advantage that the schema of datasets is extensible and multiple datasets can be mapped together. Our datasets from both sources, ClinVar and Expression Atlas, are accessed in XML formats. 

The scope of data integration models is vast, as mentioned in the literature review in the previous section. Various data integration models have been discussed by various authors including \cite{subhani2016,hamid2009} and \cite{ritchie2015}. For our study, we have adopted a meta-dimensional approach model, which refers to using multiple datasets simultaneously in the analysis \cite{ritchie2015}. This involves building a model on top of multiple datasets, which are combined or integrated either before or after building the data model. The approach facilitates the advantage of fetching information from multiple datasets and including it in the analysis model. However, the integration may also yield complex datasets resulting in less robust models. 

There are multiple methods within the meta-dimensional approach as mentioned by \cite{subhani2016} and \cite{ritchie2015}. We have adopted a concatenation-based integration method, where different matrices are combined into a large single matrix before building a model. One advantage of this method is that once it is determined how to concatenate the variables from different datasets into a single matrix, it is relatively easier to build any statistical analysis model on it. For example, on a combination of genomics datasets, \cite{fridley2012} used a Bayesian model to predict phenotypes, and \cite{mankoo2011} used Cox Lasso model to predict time to recurrence. 

It may be important to mention here that the integration attempt in this paper is only at the data level. Since the data being retrieved from public repositories is in XML format, we do not need to pre-build a structure to store data, and we are not dealing with databases either. Therefore, this method provides the advantage to avoid the data structure and storage issues. Hence, the data integration here must not be confused with the traditional database level data integration.

\begin{table}[htbp]
\caption{GENE EXPRESSION DATASET.}
\label{table 2}
\centering
\begin{small}
\begin{sc}
\begin{tabular}{lcccr}
\hline
Gene & GSM452573 & GSM452571 & GSM452642 & ... \\
\hline
AKAP9 & 3.563587736 & 3.45243272 & 3.535150355 & ... \\
AKT1 & 10.8863402 & 10.34918494 & 9.129441853 & ... \\
AKT2 & 5.005896122 & 4.463927997 &4.993673626 & ... \\
\hline
\end{tabular}
\end{sc}
\end{small}
\end{table}

\begin{table*}[htbp]
\caption{INTEGRATED DATASET.}
\label{table 3}
\centering
\begin{small}
\begin{sc}
\begin{tabular}{p{0.09\linewidth}p{0.09\linewidth}p{0.1\linewidth}p{0.07\linewidth}p{0.07\linewidth}p{0.07\linewidth}p{0.07\linewidth}p{0.095\linewidth}p{0.095\linewidth}}
\hline
Disease & Clinical Significance & Chromosome No. & Location & Variation ID & Allele ID & Gene & GSM452573 & GSM452571   \\
\hline
Long QT syndrome & Benign/ Likely benign &7 & 92001306 & 136347 & 140050 & AKAP & 3.563587736 & 3.45243272  \\
Colorectal Neoplasms & Likely pathogenic & 19 & 40236313 & 376039 & 362918 & AKT2 & 5.005896122 & 4.463927997  \\
... & ... & ... & ... & ... & ... & ... & ... & ...\\
\hline
\end{tabular}
\end{sc}
\end{small}
\end{table*}

\section{Data Integration Model}

We have used completely anonymised clinical and genomics datasets obtained from public sources. The clinical dataset (ds1) has been obtained from ClinVar \cite{ncbi}, which is an open source database that contains information about the genomic variation and links it with phenotype information. For each gene, it provides the diseases it causes and their clinical significance. In addition, it also includes the whereabouts of the gene, such as, chromosome number, location, variation ID etc. A snapshot of the data is illustrated in Table \ref{table 1}. The database was searched for 'colorectal cancer', and all the search results were downloaded and saved as XML files.

The genomics dataset (ds2) is a Gene Expression dataset of primary colorectal tumours (E-GEOD-18105), obtained from the Expression Atlas of European Bioinformatics Institute (EBI), which is a public resource for gene expression datasets \cite{gene}. Gene expression data, as the name indicates, contains information for the expression of gene(s) in a particular biological sample(s). The expression data is obtained via microarray technology, which provides parallel processing and monitoring of tens and thousands of genes, producing tons of valuable data \cite{raghavachari2013}. A typical gene expression dataset contains a matrix with genes in rows and samples in columns. The number in each cell of the matrix characterises the expression level of a specific gene in the given sample \cite{brazma2000}. Table \ref{table 2} shows an example of how gene expression data looks like. After the first column, which is gene name, the rest of the columns represents samples, and the values represent the expression levels. 

The primary reason for selecting these two datasets was that they fulfill the information requirement for this study. The ClinVar data provides the information about clinical condition against each gene present in the dataset. It also provides the clinical significance of these conditions \cite{landrum2013}. The gene expression data brings the information about the activity of those genes in different samples. Hence, the two datasets provide the required information to create an integrated dataset for this model.

\begin{table}[htbp]
\caption{STATISTICS OF INTEGRATED DATASET.}
\label{table 4}
\centering
\begin{small}
\begin{tabular}{lccr}
\hline
Output Classes & I & II \\ %& I & I \& II  \\
\hline
Unique Classes & 80 & 76 \\ %& 80  & 80 \& 76 \\ 
Feature set & 117 & 117 \\ %6 & 112 & 117  \\
Training Examples & 258 & 281 \\ % & 258 & 258 \& 281 \\
\hline
\end{tabular}
\end{small}
\end{table}

As mentioned previously, we are using a meta-dimensional approach based integration, and specifically the concatenation method. The datasets were concatenated via gene names. It has to be noted that there were multiple examples for each gene in both datasets. The examples in the ds1 with no feature sets available in ds2 were removed. On the contrary, the examples in the ds2 for which there were no feature sets in ds1, the data was extrapolated in ds1 so that the examples for that gene can be increased. Since each parameter in the feature set is independent, therefore, extrapolating some points does not affect the accuracy. 

Table \ref{table 3} shows an example of the integrated dataset, where the clinical and genomics parameters are concatenated via gene names. The statistics of the dataset is shown in Table \ref{table 4}. The data was trained with two different output classes: genes (class-I) and diseases (class-II). There are 80 unique genes, and 76 unique diseases in the dataset after removing the outliers. 

It can be argued that predicting genes as output class does not provide much meaning. Predicting disease has a more clinical value since this information is not available in the gene expression data. The reason behind this selection is only to provide an example that the classifier can be used to predict any feature from an integrated dataset without any restriction.

The resulting schema includes clinical and genomics parameters in columns, while each row represents a gene. Hence, each row tells the possible medical condition for a gene if it is active in a sample. This schema is completely flexible and scalable. It can be expanded by adding data from different sources, as long as the new data can be mapped to existing schema. More data brings more information that will only help to improve the performance of the classifier by increasing the feature set and the training examples. 

\section{Prediction Model}

In this section, we will talk about the  multiclass classification challenges, followed by the details of our prediction model, comprising the algorithm and the experimental environment.

\subsection{Multiclass Classification}\label{AA}

When we talk about disease classification, we are talking about a complicated multiclass classification problem. From classification perspective, it is relatively easier to classify binary problems or even few classes, but with increasing number of classes, the complexity of the dataset gets very high \cite{zhang2004}. The data under consideration in this study contains more than 75 different classes. When the number of output classes is that high, the variance in the data is very high as well. In such a case, it is best to have as much data as possible so that every class has a sufficient representation in training data. This is a minor limitation in our study because of the limited number of examples available from public datasets. 

\begin{figure}[htbp]
\centering
\includegraphics[width=\linewidth]{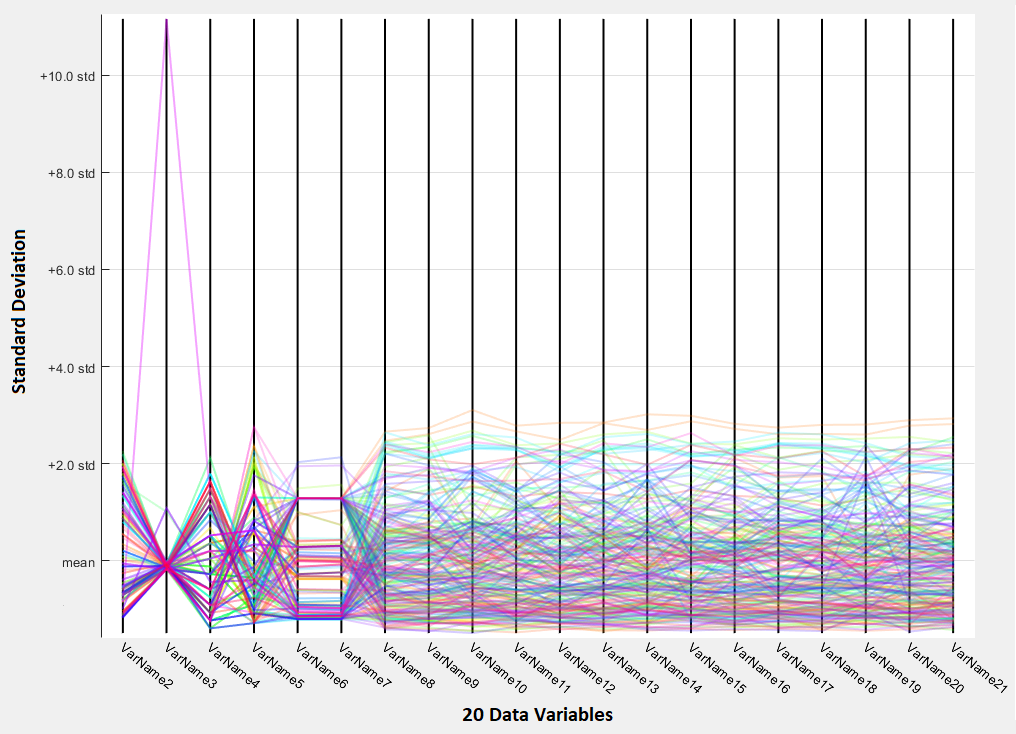}
\caption{Distribution of variance in the integrated dataset.}
\label{fig1}
\end{figure}

There is no single classification method that can be suggested to be best suited for multiclass classification \cite{zhang2004}. Any algorithm can perform better than the rest based on the characteristics and properties of the data. In this study we have used the k-Nearest Neighbours (kNN) algorithm. The reason for selecting the kNN instead of Support Vector Machines (SVM), which is a more popular classification algorithm, is our large number of output classes and the random distribution of data (Figure \ref{fig1}). Unlike SVM, which uses kernels for optimization, kNN determines the label for a given data point based on nearest data points on the distance metric. Since kNN is a non-parametric algorithm, it does not assume any explicit functions for the input data (such as Gaussian) \cite{duda1973}. This works well in our case when the data has no particular distribution and is widespread (Figure  \ref{fig1}). Hence, we can avoid the algorithmic complexity by using an algorithm which uses local optimization only. Also, kNN performs well on small to medium sized datasets \cite{duda1973}. 

\subsection{Classification Algorithm}
The kNN is a non-parametric supervised learning algorithm \cite{duda1973}. For a given dataset $X$, with labels $Y$, the algorithm calculates the distances between a new data point $z$ and all data points in $X$ to create a distance matrix. Euclidean distance is the most common method for calculating this distance. Euclidean distance between point $x_i$ and $y_i$ can be calculated by:

\begin{equation}
\label{eq1}
D(x,y)= \sum_{i=1}^k(x_i- y_i)^2  
\end{equation}

Let $R = (X_i, Y_i)$, where $i = 1, 2... N$, be the training set, where $X_i$ is the $p*q$ feature vector, and $Y_i$ is the $q$-dimensional vector which represents $m$ output class labels, as we are considering multiclass classification problem. We presume that the training data has random numeric variables with unknown distribution. 
   
From the training set $R$, the kNN algorithm narrows down to a local sub-region $r(x)$ of the input space, which is centered on an estimation point $x$. This predicting sub-region $r(x)$ contains the training points $(x')$ nearest to $x$, which can be expressed as:

\begin{equation}
r(x) = \{x'\mid D(x,x')\leq d(k)\}
\end{equation}

where, $D(x,x')$ is the distance metric between $x'$ and $x$, and $d(k)$ is the $k^{th}$ order statistic. $k[y]$ denotes the $k$ samples in the sub-region $r(x)$, which are labelled $y$. The kNN algorithm estimates the posterior probability $p(y \mid x)$ of the estimation point $x$:

\begin{equation}
p(y \mid x) =  \frac{p(x \mid y)p(y)}{p(x)} \cong \frac{k[y]}{k}
\end{equation}

Generally, when the kNN is used for binary classification, the label assignment is relatively easier since the algorithm has to select between two classes only, such as : 

\begin{equation}
g(x)=
\begin{cases}
1, k[y=1] \geq k[y=-1] \\
-1, k[y=1] \leq k[y=-1]
\end{cases}
\end{equation}

We have improvised this functionality for our study, where the output class is non-binary. In this case, for any estimation point $x$, the decision $g(x)$ for a given label $y$ is estimated by:

\begin{equation}
g_k (x)=y_k \mid min D_k  
\end{equation}

where, $D_k$ is represented by \ref{eq1} Hence, the decision that will maximise the posterior probability will be assigned for the output label. For a multiclass classification problem, where $y \in \{1 \dots k\}$, the kNN algorithm uses the following decision rule: 

\begin{equation}
F(x) = argmax [g_k (x)]
\end{equation}

Thus, for the selected nearest $k$ neighbours, the algorithm calculates the posterior probability for each class, and the class with highest probability is assigned to $x$. Euclidean distance is the most common method, but there are other distance calculation methods as well, such as seuclidean, mahalanobis, spearman, etc \cite{duda1973}. 

\subsection{Performance Measurement}

Generally, the performance of machine learning classifiers is measured using various parameters, such as, accuracy, sensitivity, specificity, and Receiver Operator Curve (ROC). These parameters are calculated based on the true positives, true negatives, false positives and false negatives of classifier. For binary classes, these parameters are easier to calculate because there is only one positive and one negative class. However, for multicalss classification, the problem is more complicated and it is not easy to calculate each parameter for each class. Especially ROC, which is a standard measure to represent performance of a classifier, is very complicated to calculate for a very large multiclass problem. This problem has been discussed in further detail by Fawcett in \cite{fawcett2006}.

Therefore, calculating each parameter for every class will not only be laborious, but will also produce loads of results that will be difficult to ensemble and explain. To simplify that, we have only used confusion matrices to represent the performance of the classifier and used the accuracy for each classifier to compare the results for the two classes.

\subsection{Experimental Environment}

We have used Matlab (R2018a) for all the experiments, which provides built-in libraries for machine learning classifiers. We used the machine learning toolbox to train the classification model using kNN. The toolbox takes the data as input and process the classification itself using the built-in library functions and selected features. The classification toolbox uses the Euclidean distance by default to compute the distance metrics. The tool box can be used to reproduce the results.

At first, we perform the Principal Component Analysis (PCA) for dimensionality reduction. Since, our data is multivariate, ranging from gene expression data to phenotypic data, the data points are widespread in the data space. Figure \ref{fig1} shows the standard deviation distribution of the first 20 data variables from the integrated dataset. It can be seen that the data distribution is very random and does not follow any standard distribution function. Therefore, it is important to reduce the dimension of the integrated dataset. We performed PCA to explain 95\% variance in the data.

\begin{figure}[h]
    \centering
    \includegraphics[width=\linewidth]{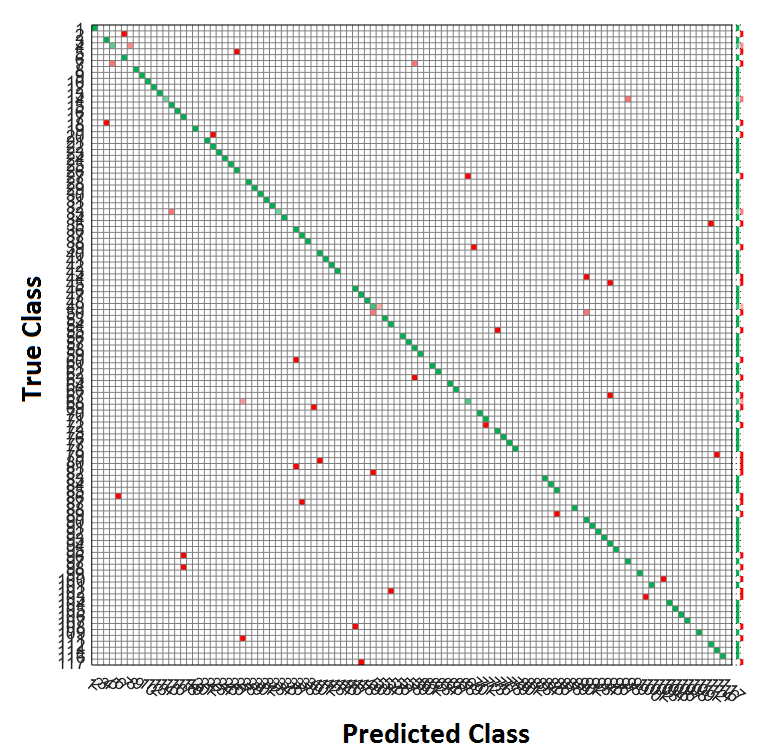}
    \caption{Confusion matrix for class-I.}
    \label{fig2}
\end{figure}

The results of classification depend highly on the dimensions of the dataset. The correlation between the number of examples and feature sets is very critical in this case to avoid over-fitting \cite{hua2004}\cite{jain1978}. The clinical dataset has only 5 features, which is not a large enough set to be used stand-alone for prediction model. With a feature set of 5, the prediction is neither reliable nor comparable with other datasets. The genomics dataset is large enough in this respect, but it does not contain the class-II so we cannot predict diseases. Therefore, we have only used the integrated dataset to train with the prediction model explained in the previous section (IV), and then compared it with other classifiers.

The results are validated using 10-fold cross-validation. This means, the dataset is divided into 10 parts; one part is held out as a test data and the rest of the 9 parts are used as training data. This step is repeated 10 times using a different part every time to holdout as a test data. This way every example from data is used both as training and test data. The resulting accuracy is an average of the 10-fold process. 

\section{Results}

The performance of a classification model is analysed using a confusion matrix. Figure \ref{fig2} shows a confusion matrix for class-I prediction. The rows in a confusion matrix represent the true output class, and the columns represent the predicted class. The diagonal cells indicate the true positives (green) and the false negatives; and the off-diagonal cells indicate the false positives and the true negatives (red). The bottom right cell shows the overall accuracy and the loss of the classifier.

\subsection{Classification with our Classifier}

The number of neighbours (NN) is a variable in the algorithm, which can be tuned to change the performance of the algorithm. We tested the performance of the algorithm over 10 different neighbours, from 1 to 10.

\begin{figure}[htbp]
\centering
\includegraphics[width=\linewidth]{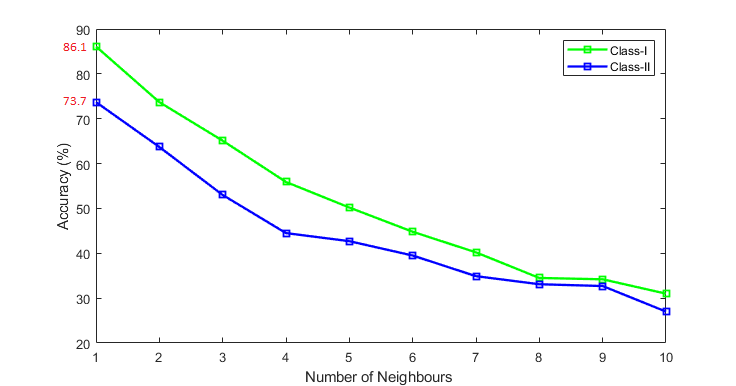}
\caption{Accuracy of classification model for both classes.}
\label{fig3}
\end{figure}

As mentioned previously, we trained the integrated dataset for two different classes: genes (class-I) and diseases (class-II). The results are shown in Figure \ref{fig3}. At NN=1, the trained model predicts class-I with 86\% accuracy, and class-II with 73\% accuracy. 

Initially, the accuracy drops almost linearly with the increasing number of neighbours. The drop in accuracy can be attributed to the variation in data. As the algorithm considers more number of neighbours, each neighbour brings more variation that affects the prediction accuracy. However, as it can be seen in Figure \ref{fig3}, accuracy remains above 50\% for up to 3 neighbours for both classes which can be regarded as a good accuracy considering a multivariate training data. Following NN=4, the accuracy drops almost exponentially.  

This variation over neighbours may be avoided by introducing a weighted parameter in the algorithm. This parameter weighs the contribution of each neighbour under consideration based on its distance. The nearest neighbours gets higher weights than the distant ones. Matlab's classification tool uses the squared inverse method to calculate the weights, which can be expressed as:

\begin{equation}
w_n=  \frac{1}{d(x_n-x_i)^2}
\end{equation}

where, $x_n$ is the neighbour to point $x_i$. To accommodate this weight parameter, the eq. \ref{eq1} is adjusted as follows:

\begin{equation}
D(x,y)= \sum_{i=1}^k w_i(x_i- y_i)^2 
\end{equation}

\begin{figure}[htbp]
\centering
\includegraphics[width=\linewidth]{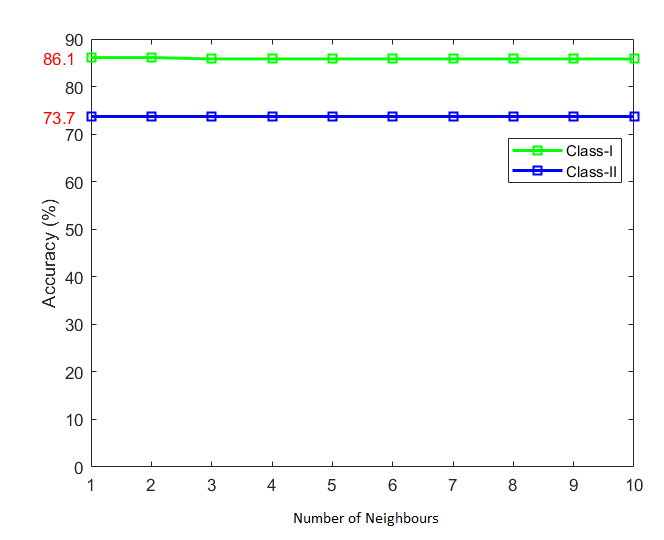}
\caption{Accuracy for both classes with weighted kNN.}
\label{fig4}
\end{figure}

We tested this updated version by training the integrated dataset, and we observed that the accuracy was raised to the maximum (86.1\% for class-I and 73.7\% for class-II) for all NN's. The results are shown in Figure \ref{fig4}. This is perhaps because the weighted version predicts based on the neighbour with the highest weight. Since the nearest neighbour is most likely to have highest weight out of all neighbours, the classification result is the same every time. This result seems to be not very helpful for our dataset.

However, our model has predicted the diseases with up to 73\% accuracy. The accuracy is not as good as for the class-I (86\%). There can be multiple reasons behind this. The representation of each class label in the data varies, which affects the prediction accuracy. Some classes have sufficient examples in the data, while others have only few examples. The higher the representation of a class label in the training data, the better is the prediction accuracy for that class. The distribution of class-I labels in dataset is comparatively more uniform than class-II; hence, higher accuracy. Still, achieving 73\% accuracy for class-II is a very good result considering the size, shape, and multivariate nature of the dataset.

\subsection{Comparing with other Classifiers}

\begin{figure}[htbp]
\centering
\includegraphics[width=\linewidth]{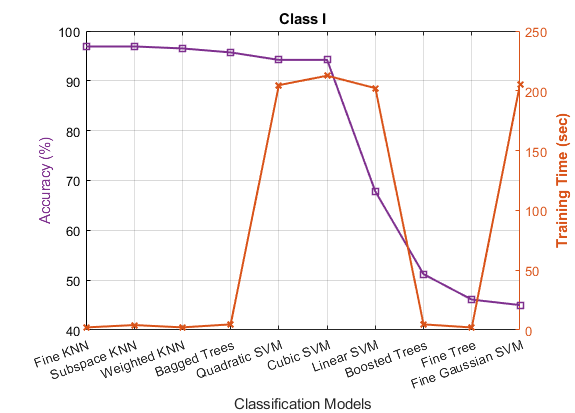}
\caption{Performance of other classification models for class-I predictions.}
\label{fig5}
\end{figure}

We trained the same integrated dataset with other classifiers in order to compare the performance. Using PCA of 95\%, we trained all the classifiers available on Matlab's classification toolbox, and then selected the top 10 models (out of 22) to compare the classification accuracy for both classes. NN=1 for all the models in the classification toolbox. 10-fold cross-validation was used to avoid over-fitting. The results are shown in Figure \ref{fig5}.

For class-I, the kNN models provided the highest accuracy of 96.9\%. kNN was followed by the Tree and SVM models. As we can see, the top three models are all kNN models providing accuracy of above 95\%. The accuracy of the SVM models (Quadratic and Cubic) is almost in the same range (95-96\%), however, the training time for the SVM models is 200 times higher than the kNN models. This is because the SVM uses the cost minimization functions, such as gradient descent or kernel functions, which take much longer to converge. Since kNN does not use any of those functions, it is more robust and provides with the same, rather better accuracy. To summarise, although both kNN and SVM models have predicted about the same accuracy, the kNN models are much more robust than the SVM models in terms of performance.

The tree models, except for bagged trees, performed poorly providing accuracy of about 50\% or under. The training time of the tree models is as good as that of kNN models (few seconds), but the accuracy is poor. Bagged trees, which is a bootstrapping method, performed quite well. On the other hand, boosted trees provides an accuracy of just about 51\%. Although both of them are ensemble methods, which means they provide an average of multiple models trained on a subset of data, bagged trees provided much better result. 

The accuracy of these kNN models (Fine kNN, Subspace kNN, and Weighted kNN) is slightly higher than our prediction model (Figure \ref{fig3}). The reason for this is that the models in the toolbox are set on different defaults and use different functions than the ones we used. The classification function that we used is primarily for multiclass classification problems. On the other hand, the function used by the toolbox models are mainly designed for binary problems, hence, the difference of accuracy.

\begin{figure}[htbp]
\centering
\includegraphics[width=\linewidth]{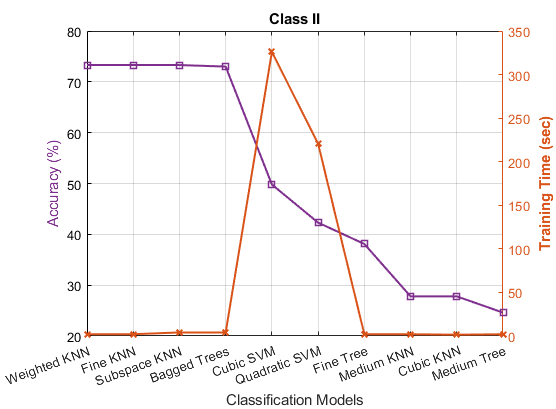}
\caption{Performance of other classification models for class-II predictions.}
\label{fig6}
\end{figure}

Similar results are seen for class-II. The results are shown in Figure \ref{fig6}. The top 10 models selected here are slightly different than those for class-I, but majority are the same. The highest accuracy achieved for class-II is 73.3\%, which is just about the same as achieved by our model (Figure \ref{fig3}). The top 3 models are all kNN models, with bagged trees standing at 4th position with 73\% accuracy. All SVM models provide accuracy of less than 50\% with training times as high as over 200 times of the kNN models. The same is the case for tree models except for bagged trees; same result as for class-I. A plausible explanation for good performance of bagged trees could be that they perform better on high dimensional data.

\section{Discussion}

We have demonstrated a novel way for multiclass classification based on integrated clinical and genomics datasets. We have used concatenation-based data integration model for this purpose, which has been discussed by various researchers before (\cite{subhani2016}\cite{ritchie2015}), but not implemented in the area of health care. Hence, this is the first time that we have attempted to use this meta-dimensional approach to integrate datasets.

In the past, people have used various other methods for data integration such as tree-based models \cite{nevins2003}, statistical models \cite{lee2009}, and graph based models \cite{kim2014}\cite{schadt2005}\cite{yang2014}. All these models require considerable amount of effort and time to build the data models first, before creating the data analysis model, such as building the binary trees, or creating graphs models from datasets. Our method does not involve any of those complex models; it only requires concatenation of all the datasets into a single matrix. Once concatenated, the model transfers the dataset directly to the analysis model and starts training the learning algorithm. Hence, it is way more efficient in terms of time and computational costs as compared to other methods.

In terms of analysis, from our knowledge, none of the previous models have been used for multiclass disease classification problems in health care. They have only been demonstrated for binary classifications; and, therefore, their results cannot be compared with our model, which is a multiclass classification model. 

In terms of data models, it will be very difficult to perform multiclass classification based on the previously mentioned models because they will require to build a separate data model (trees of graphs) for each output class before the analysis model. Having multiple output classes, the analysis models will get extremely complicated with several input data models. With our proposed model, as there is only single concatenated dataset, the multiclass classification is less complicated and manageable because the dataset has only one data model with a single schema.  

Since, we could not compare our results with any other previous results from other researchers, we have demonstrated comparison with other classification models. The results shown in Figures \ref{fig5} and \ref{fig6} demonstrate that the kNN models can outperform the rest of the classification models in terms of prediction accuracy and performance.

Our proposed approach provides a very flexible and scalable model, along the lines of our previous work as reported in \cite{anjum2006,van2005,van2005grid,kiani2013}, which can be scaled to adjust any new dataset and accommodate any analysis model. As long as there is a relational dataset, it can be concatenated to the existing dataset within the same data model and schema. Any analysis model or algorithm, including prediction, classification, regression models, can be built on top of the dataset. This flexibility enables this approach to be adapted for any research purpose in any domain. 

\section{Conclusion and Future Directions}

The way forward in precision medicine is to use all available data from clinical and genomics domains in order to provide the best clinical solutions. The datasets need to be intelligently integrated for this purpose. In this paper, we have performed clinical predictions based on clinical and genomics information. We have attempted to integrate a clinical (ClinVar) and a genomic (gene expression) dataset, and performed classification for disease predictions. We have designed a multiclass classification model that predicts diseases from integrated datasets. The model, which is validated by 10-fold cross-validation, has predicted diseases with up to 73\% accuracy. We also predicted genes as an extra variable, from the same dataset, and achieved up to 86\% accuracy. We have compared the results with other classification models and demonstrated that our model outperforms the rest. We can conclude that constructing the learning classifiers on top of large-scale inter-domain integrated datasets can provide very good clinical predictions. This can prove to be very beneficial and a stepping-stone towards the precision medicine. 

This research study shows that diseases can be predicted with good accuracy from a patient's dataset if it has both clinical and genomics parameters present. The accuracy will further improve if we train the model with a much larger size of training data. The reliability and confidence in results will increase by incorporating more clinical and genomics information. We have demonstrated with a gene prediction example, that, when the dataset is more uniformly distributed among different classes, the prediction accuracy goes high even on a multiclass classification task.

This study has great potential to expand including achieving analysis provenance \cite{mcclatchey2013}. The more information a dataset will contain, higher the accuracy can be achieved. The dataset can be expanded to include more multivariate clinical and genomics datasets, such as clinical trials and multi-omics datasets, respectively. Including clinical information from clinical trials or laboratory tests will have a significant impact in the clinical prediction studies.

\section*{Acknowledgment}

The authors would like to thank and acknowledge the help and support received from Usman Yaseen, Sanna Aizad, Bilal Arshad and Craig Bower in the successful completion of this study. 

\bibliographystyle{IEEEtran}
\bibliography{IEEEabrv,bibtemplate}

\end{document}